\documentclass[twocolumn]{aastex63}
\usepackage{bm}
\usepackage{threeparttable}
\usepackage{morefloats}
\usepackage{CJK}

\def\bfnabla{{\mbox{\boldmath $\nabla$}}}

\def\msun{M_\odot}
\def\mbh{M_{\rm{BH}}}
\def\Medd{\dot{M}_{\rm{Edd}}}
\def\Ledd{L_{\rm{Edd}}}

\renewcommand\bv{{\mbox{\boldmath $v$}}}

\def\<{\,\langle\langle}
\def\>{\,\rangle\rangle}

\shorttitle{Convection in AGN Disk.}
\shortauthors{Jiang \& Blaes}

\begin{document}
\begin{CJK*}{UTF8}{gbsn}

\title{Opacity Driven Convection and Variability in Accretion Disks around Supermassive Black Holes}

\correspondingauthor{Yan-Fei Jiang}
\email{yjiang@flatironinstitute.org}

\author[0000-0002-2624-3399]{Yan-Fei Jiang(姜燕飞)}
\affiliation{Center for Computational Astrophysics, \\
Flatiron Institute, \\
New York, NY 10010, USA}

\author{Omer Blaes}
\affiliation{Department of Physics, \\
University of California, \\
Santa Barbara, CA 93106, USA}

\begin{abstract}
We study the structure of accretion disks around supermassive black holes in the radial range  $30\sim 100$ gravitational radii, using a three dimensional radiation magneto-hydrodynamic simulation.  
For typical conditions in this region of Active Galactic Nuclei (AGN), the Rosseland mean opacity is expected to be larger than the electron scattering value. We show that the iron opacity bump causes the disk 
to be convective unstable. Turbulence generated by convection puffs up the disk due to additional turbulent pressure support and enhances 
the local angular momentum transport. This also results in strong fluctuations in surface density and heating of the disk. The opacity drops with increasing temperature and 
convection is suppressed. The disk cools down and the whole process repeats again. This causes strong oscillations of the disk scale height and luminosity variations by more than a factor of $\approx 3-6$ over a few years' timescale. Since the iron opacity bump will move to different locations of the disk for black holes with different masses and accretion rates, we suggest that this is a physical mechanism that can explain the variability of AGN with a wide range of amplitudes over a time scale of years to decades.  

%The simulation results may explain the formation of X-ray coronae in Active %Galactic Nuclei (AGNs), the compact size of such coronae, and the observed
%trend of optical to X-ray luminosity with Eddington ratio for many AGNs. 

\end{abstract}

\keywords{accretion, accretion disks --- (galaxies:) quasars: supermassive black holes --- 
                magnetohydrodynamics (MHD) --- methods: numerical ---  radiative transfer}

\section{Introduction}

The standard model of geometrically thin, radiatively efficient accretion disks
\citep{SS73,NOV73} provides a good, observationally tested,
first-order model for the soft states of X-ray binaries (both black hole and
neutron star, e.g. \citealt{DON07}).  Accounting for proper opacities
and the effects of irradiation, it has also been developed into a testable
model that successfully explains the outbursts observed in dwarf novae and
low mass X-ray binaries \citep{LAS01}.  However, its application to bright
active galactic nuclei (AGN) and quasars, where the model still predicts
the existence of a geometrically thin disk, has always been problematic
(e.g. \citealt{KOR99,ANT13}).  Beyond the most basic prediction of thermal
emission in the ultraviolet \citep{SHI78,MAL83}, the model generally fails
to provide a good description of the far ultraviolet spectral energy
distribution (e.g. \citealt{SHA05,LAO14}).  While it successfully predicts that shorter wavelength
optical/ultraviolet photons originate from smaller distances from the black
hole than longer wavelength photons, consistent with the measured trends in continuum reverberation mapping \citep{EDE15} and microlensing campaigns \citep{Morganetal2010,Blackburneetal2011,Mosqueraetal2013},
it is {\it not} consistent with the absolute emission region sizes.

One of the most glaring observational discrepancies is the extreme
variability observed in what are now called changing look quasars (e.g.
\citealt{LAW18} and references therein).  Optical emission from these
sources are observed to increase and/or decrease by factors of two to ten
on time-scales of a few years to  a decade (e.g. \citealt{LAM15,MacLeodetal2016,RUA16,Yangetal2018,Dexteretal2019}), accompanied by
a resurgence and/or loss of the broad emission lines. These changing look AGNs appear to be the tail of a continuous distribution of quasar properties where the large amplitude 
variability is likely caused by physical processes in the disk \citep{Rumbaughetal2018,Luoetal2020}.
However, this variability is
far shorter than the ``viscous" time scale (required to move mass radially from the outer edge of the disk to the inner region)
of the standard disk model, but is not inconsistent with the thermal time
scale (required to heat or cool the disk).  These two time scales differ
by a factor of the square of the ratio of the radius $R$ to the local disk
scale height $H$.  One possible resolution is therefore to suppose that the
actual accretion flow is geometrically thick $H/R\sim1$ in the optically
emitting regions, perhaps because of vertical support by magnetic fields
\citep{DEX19}.

The fact that the standard model of accretion disks does such a poor job
of explaining observations suggests that one should look at ways in which
AGN accretion disks necessarily differ from their counterparts in X-ray
binaries and cataclysmic variables.  One very important difference is that
accretion disks in bright AGN have thermal pressures which are hugely
dominated by radiation pressure.  Moreover, their ultraviolet temperatures
mean that they are subject to large opacities, the effects of which are
to make radiation pressure forces even stronger.  One important aspect of
this is the likely presence of line-driven outflows from the disk
\citep{PRO00,LAO14}.

But even within the Rosseland photosphere of the disk,
opacity effects can be extremely important.  Sophisticated one-dimensional
models of the vertical structure of the disk generally exhibit density
inversions due to enhancements of Rosseland opacity with declining outward
temperature (see, e.g., Figure 9 of \citealt{HUB00}).  Such density inversions
are also commonly seen in one dimensional models of radiation pressure
supported massive star envelopes \citep{JOS73,PAX13}, which have similar
density and temperature conditions to AGN disks.
The density inversion can be either driven by the opacity peak due to 
hydrogen and helium ionization fronts as studied by \citet{HUB00}, or the iron 
opacity bump around the temperature $1.8\times 10^5$ K \citep{JIA18}.
In this paper, we focus on the hotter opacity bumps due to irons. 
%It is generally the iron
%opacity bump which drives such density inversions\footnote{The non-LTE models of \citet{HUB00} included only hydrogen and helium, with %no metals.  The vertical density inversions they found arose from vertical ionization fronts of hydrogen and helium, not the iron %opacity bump explored here in this paper.  Nevertheless, such inversions would also be expected to drive strong convection with all the %consequences that we have found here.  {\color{red} Perhaps move this to the discussion?}},
%Unfortunately, the Hubeny et al. (2001) models that did include metals did not discuss the vertical density structure.  Because they %only included continuum, not line, opacities, I suspect that they did not really have an iron bump, but it's hard to say.

Density inversions due to the opacity peaks are of course
unstable,
and simulations of these inversions show considerable convective turbulence
\citep{JIA15}.  However, under conditions of moderate optical depth (optical
depth per pressure scale height less than the ratio of the speed of light
to the gas sounds speed), convective heat transport is inefficient and leads
to a porous structure with large density fluctuations associated with shocks.
Moreover, the time-averaged structure still maintains a vertical density
inversion.  The large density fluctuations can also trigger even larger
enhancements in opacity due to helium recombination, possibly triggering
bursty outflows in massive stars \citep{JIA18}. AGN disks are generally
expected to be in this low optical depth, rapid radiative diffusion regime,
and so similar behavior might be expected.

Accretion disks are considerably more complex compared to stars, however,
because of the coupling between the thermal state of the disk and angular
momentum transport.  The standard accretion disk model with an $\alpha$
stress prescription proportional to total thermal pressure is unstable
to thermal and ``viscous" instabilities \citep{LIG74,SHA76}.  Modern radiation
MHD simulations of MRI turbulence appear to confirm the thermal instability
\citep{JIA13,MIS16}, although it can be stabilized if the disk is largely supported by magnetic rather than thermal pressure \citep{SAD16}. 
Local shearing box simulations show that stability can also be achieved under 
AGN conditions by the iron opacity bump \citep{JIA16}.

Local shearing box simulations of MRI turbulence in white dwarf
\citep{HIR14,COL18,SCE18} and protostellar \citep{HIR15} accretion disks show
that convection can significantly enhance magnetorotational (MRI) turbulence.
It seems reasonable to expect that the opacity-driven convection from
unstable density inversions in AGN disks might also lead to interesting
variations in MRI turbulent stresses.  Large turbulent kinetic energy
densities might also provide support against the vertical tidal gravity
of the black hole similar to what is found in massive stars \citep{JIA18}.

In this paper, we report the results of an initial investigation of these
opacity effects using global radiation MHD simulations of AGN accretion disks.
This work builds on previous efforts to simulate the near-black hole regions
of AGN accretion disks, beginning with super-Eddington flows
(\citealt{JIA14,JIA19b}; see also \citealt{SN16}) and sub-Eddington,
magnetically supported flows \citep{JIA19a}.  The temperatures in these
previous simulations were all too high for the iron opacity bump to be
present, and the flux mean opacity was dominated by electron scattering.
Here, using very similar initial conditions to those in \citet{JIA19a}, we
simulate a region of the disk further out from the black hole (beyond
thirty gravitational radii) where we achieve low enough temperatures that
interesting opacity effects take place.  The simulations exhibit a high
degree of variability in luminosity, intermittent episodes of convection,
and rapid and variable radial diffusion of mass, which can all be traced
to the effects of variable opacity in these highly radiation pressure
dominated flows.

This paper is organized as follows.  In section 2 we briefly review the
simulation methods and the physics that is included, as well as the spatial
grid and initial conditions.  We provide a detailed description of the
physical behavior exhibited in the simulation in section 3.  In section 4
we discuss the potential applications of our work to explaining observations
of AGN, as well as other accretion-powered sources, and we summarize our
conclusions in section 5.

\section{Simulation Setup}
\label{sec:setup}
We solve the ideal MHD equations coupled with the time-dependent, frequency-integrated radiative transfer equation for intensities over 
discrete angles in the same way as described in \cite{JIA19b}. We use spherical polar coordinates in {\sf Athena++} (Stone et al. 2020, submitted) 
covering the radial range from $30r_g$ at the inner radial boundary to $1.2\times 10^4r_g$ at the outer radial boundary, where $r_g$ is the gravitational radius of the black hole. At each radius, the simulation domain includes the whole sphere with polar angle $\theta\in[0,\pi]$ and azimuthal angle $\phi\in[0,2\pi)$. We are most interested in the radial range around $50-100r_g$ where we anticipate that the disk will have enhanced Rosseland opacity due to the iron bump.  We therefore do not extend the simulation domain all the way down to the innermost stable circular orbit (ISCO) to avoid the very small time step that would be necessary to simulate these innermost regions.  The simulations we have done in the past covering the inner $\sim 30r_g$ do not show any interesting opacity effects because the density is too low and the temperature is too high there \citep{JIA19a}. The effects of having an inner boundary all the way to the ISCO will be studied in future investigations.

\subsection{Boundary and Initial Conditions}

We used static mesh refinement to resolve the inner disk near the midplane region. The root level has resolution $\Delta r/r=\Delta \theta=\Delta \phi=0.098$ 
and we use four levels of refinement with the finest level covering the region $(r,\theta,\phi)\in [30r_g,400r_g]\times [1.48,1.66]\times [0,2\pi)$, corresponding to a resolution of $\Delta r/r=\Delta \theta=\Delta \phi=6.1\times 10^{-3}$. We use 80 discrete angles in each cell to resolve the angular distribution of the radiation field.
The simulation is done for a black hole with mass $\mbh=5\times 10^8\msun$ located at $r=0$. We use the Pseudo-Newtonian potential \citep{PaczynskiWiita1980} $\phi=-G\mbh/(r-2r_g)$ for the black hole, which is actually pretty close to the Newtonian formula in the radial range we simulate. Here $G$ is the gravitational constant, $r_g\equiv G\mbh/c^2$ is the gravitational radius and $c$ is the speed of light.
For the inner boundary condition at $r=30r_g$, we simply allow all gas and radiation to flow inward but do not allow anything to come out of the inner boundary. In particular, we are neglecting any effects due to irradiation from the inner disk inside $30r_g$ in this simulation.  For the outer radial boundary condition at $1.2\times 10^4r_g$, we only allow gas and radiation to leave the simulation box but do not allow anything to come in. 

We initialize the simulation with a rotating torus centered at $400r_g$. The inner and outer edges of the torus are at $226r_g$ and $1050r_g$ respectively. For the entire region in which we are interested (inside $\sim 100r_g$), the density is initialized to be the floor value ($10^{-17}$ g/cm$^3$) in order to minimize the artifacts of the initial condition on the properties of the accretion disk. The torus structure is the same as we used in \cite{JIA19b}. The initial temperature at the center of the torus is $3.96\times 10^5\ \ K$ and it drops to $2\times 10^4\ \ K$ at the inner and outer edges. The ratio of radiation pressure to gas pressure varies from $150$ to $450$ inside the torus. 
We initialize the $\phi$ component of the vector potential to be proportional to density and set other components to zero. This results in a big loop of magnetic field initially in the torus, which has a locally net poloidal component through the midplane. The initial magnetic pressure near the inner edge of the torus is about $70$ times the gas pressure and $\sim 10^{-3}$ of the radiation pressure. Due to the limited simulation duration that we can afford to do, only gas near the inner edge of the torus will have enough time to flow towards the inner region and form a disk with self-consistent structure.

%Inner torus reaches r=226r_g, centered at 400r_g

\begin{figure}[htp]
\centering
	\includegraphics[width=1.0\hsize]{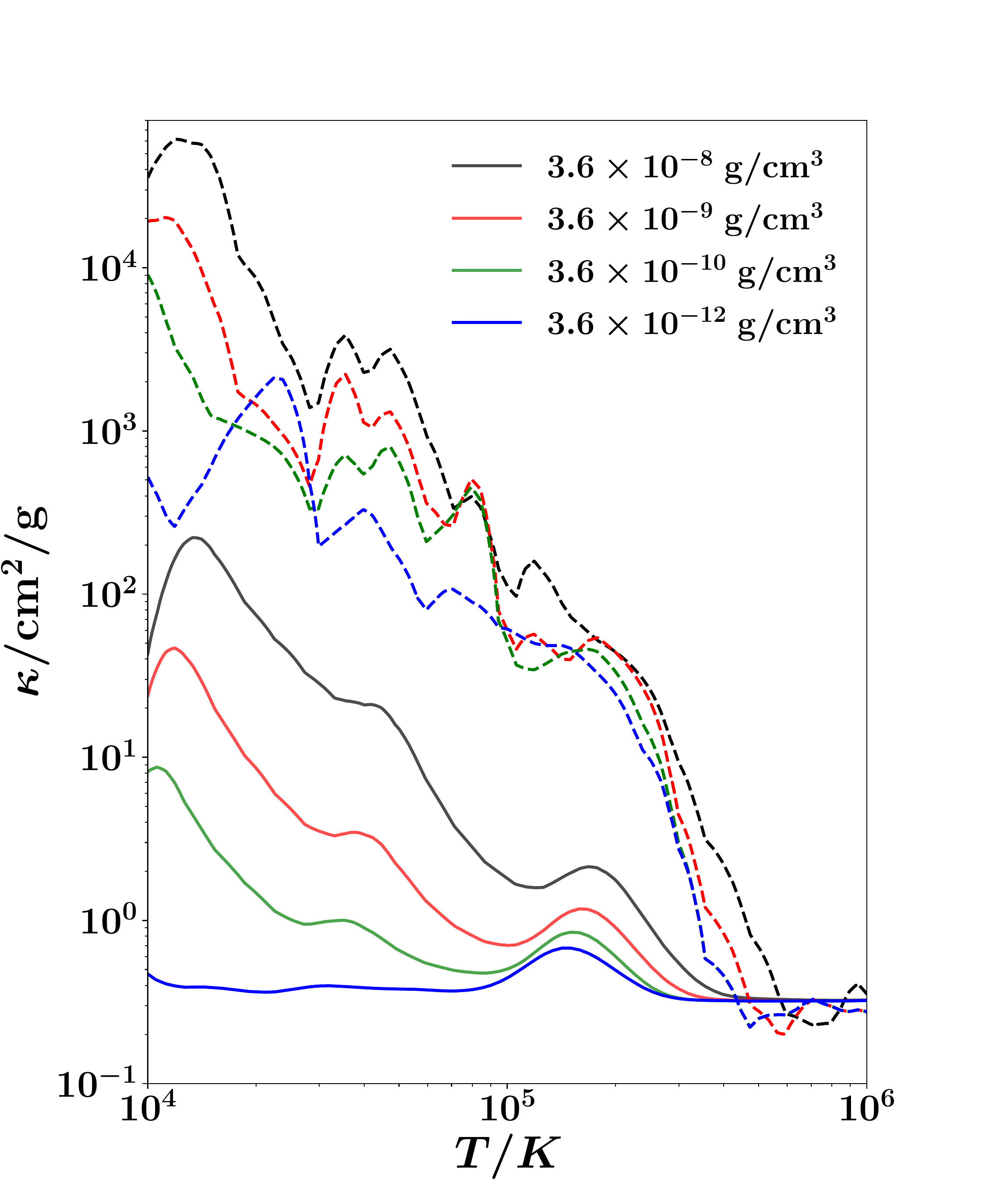}
	\caption{The Rosseland mean (solid) and Planck mean (dashed) opacities used in the simulation as a function of gas temperature for 
	four different fixed densities ranging from $3.6\times 10^{-8}$ g/cm$^3$ to $3.6\times 10^{-12}$ g/cm$^3$, as indicated in the legend.  The Planck 
	mean opacities are significantly larger than the Rosseland mean values for $T<3\times 10^5 K$. 
}
	\label{opacity}
\end{figure}

\subsection{The Opacity}
For typical temperatures and densities in accretion disks around supermassive black holes, the relevant opacity is not dominated by electron scattering 
and free-free absorption opacity as is commonly assumed in classical accretion disk models for X-ray binaries \citep{JIA16}. Instead, contributions from many lines 
 increase the effective continuum opacity significantly. Due to the changing ionization states of different species, the effective Rosseland mean and Planck mean opacities also show complicated dependencies on gas density and temperature. This is potentially interesting as it can drive hydrodynamic instabilities \citep{HEA72,BlaesSocrates2003}. To capture these opacities accurately, we adopt the OPAL opacity table\footnote{The Planck mean opacity is from the TOPS opacity at this website \url{ https://aphysics2.lanl.gov/apps/}}  \citep{iglesias96} for the Rosseland and Planck means. These opacities are shown in Figure \ref{opacity} as a function of temperature for four different densities, assuming solar metallicity.  They are very similar to the opacities of protostellar disks for $T>10^4$ K \citep{ZhuJiangStone2019} and the envelopes of massive stars \citep{JIA15}. Figure \ref{opacity} shows that the Planck mean opacity is typically orders of magnitude larger than the Rosseland mean opacity for the temperature range $10^4\sim 5\times 10^5$ K.  The enhancement in Rosseland opacity around $T=1.8\times 10^5$ K is the iron opacity bump \citep{PAX13,JIA15}, while the increase below $10^5$ K is due to ionization of hydrogen and helium. Note that the iron opacity bump is typically just a few times larger than the electron scattering value, but we will show that this can have a dramatic effect on the structure and dynamics of quasar accretion disks.  In the simulation, the opacity in each cell is calculated with bilinear interpolation of the opacity table based on local gas temperature and density.

\begin{figure*}
    \includegraphics[width=\textwidth]{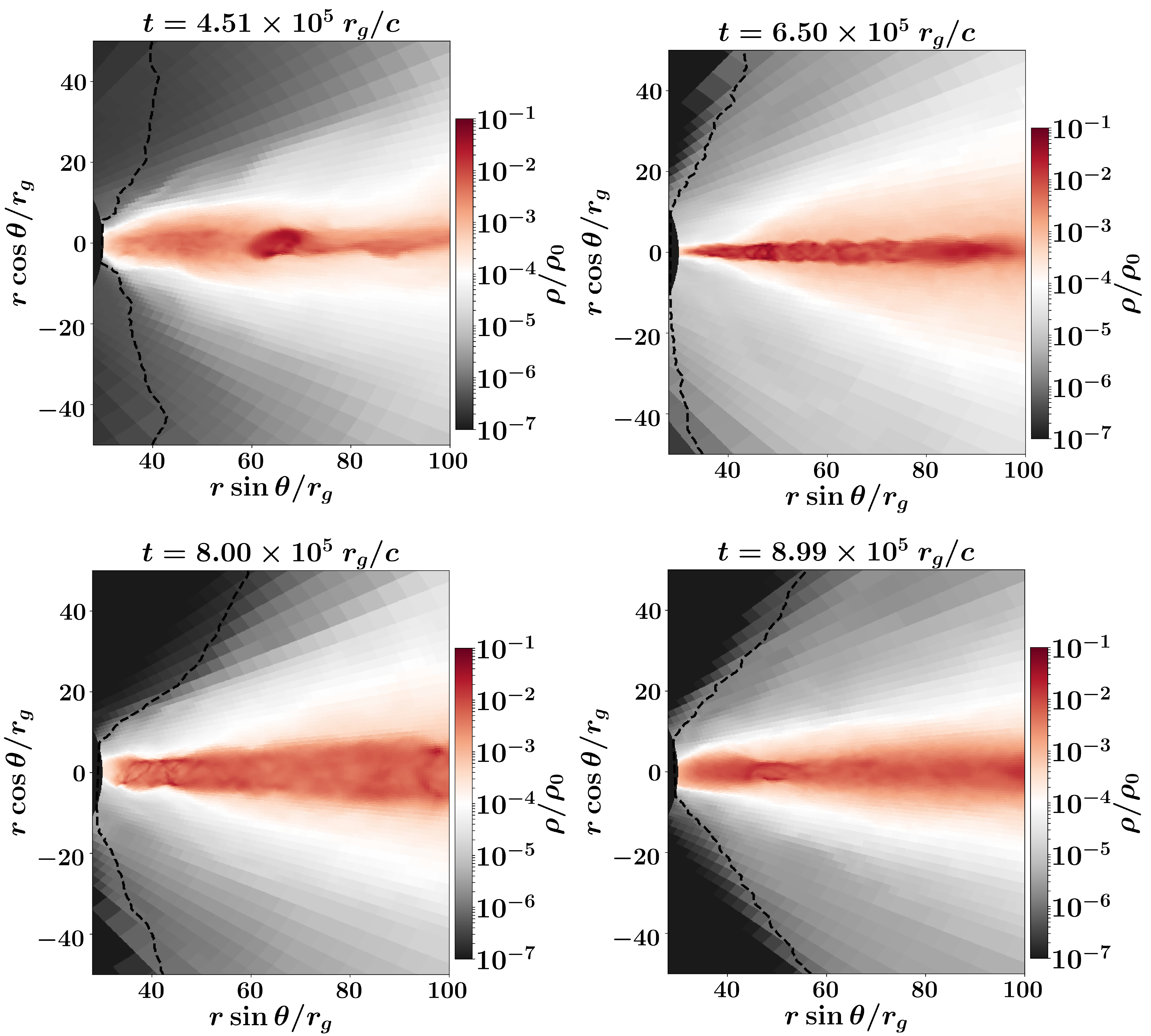}
    \caption{Poloidal distribution of azimuthally-averaged density  at four representative epochs. The dashed black lines are the location where the  integrated Rosseland mean optical depth from the rotation axis is unity. The density is scaled with the fiducial value $\rho_0=10^{-8}\  {\rm g/cm}^3$.
	}
    \label{fig:rhov2Dsnapshots}
\end{figure*}

\begin{figure*}
    \includegraphics[width=\textwidth]{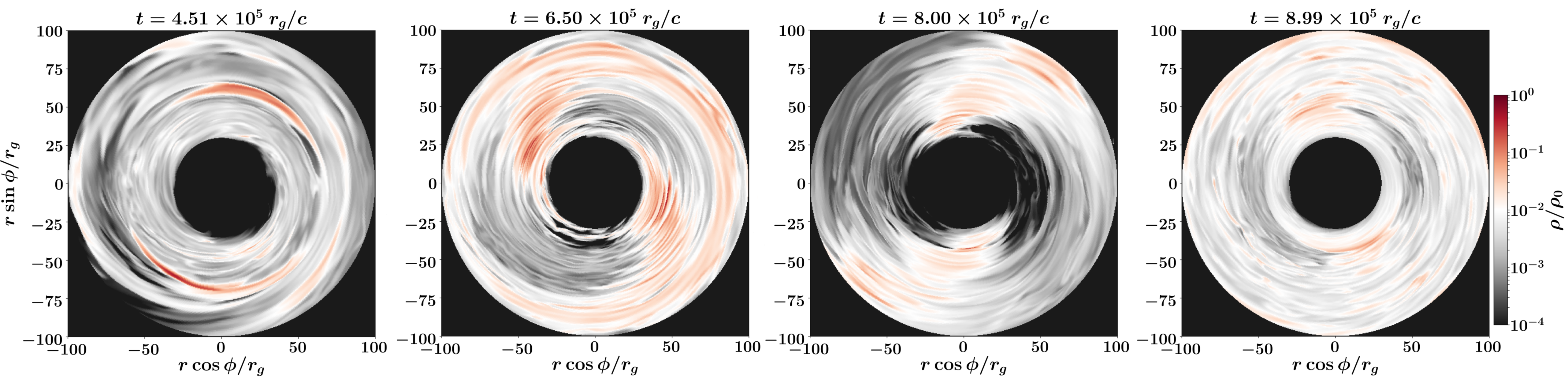}
    \includegraphics[width=\textwidth]{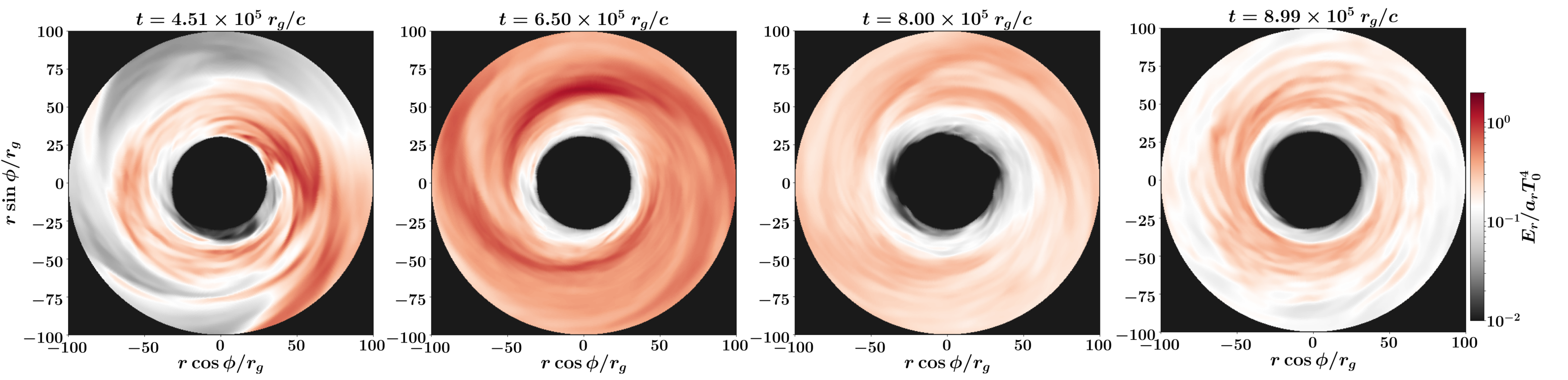}
    \caption{Density (top panels) and radiation energy density (bottom panels) distributions in the equatorial plane out to radius $100r_g$ at the same four representative epochs as shown in Figure~\ref{fig:rhov2Dsnapshots}. The density is scaled with the same fiducial value $\rho_0$ while the radiation energy density is scaled with $a_rT_0^4$ with the fiducial temperature $T_0=2\times 10^5$ K. 
        }
    \label{fig:rhomidsnapshots}
\end{figure*}

\section{Results}
\label{sec:result}
We will describe our simulation results using the following set of fiducial numbers to scale all the quantities: 
density $\rho_0=1.0\times 10^{-8} ~{\rm g/cm}^3$, temperature $T_0=2\times 10^5$ K, gas pressure and energy densities $P_0=2.77\times 10^5~{\rm dyn/cm}^2$, length $r_g=7.42\times 10^{13}~{\rm cm}$ and velocity $v_0=5.26\times 10^6~{\rm cm/s}$. The fiducial time unit is $r_g/c=7.8\times 10^{-5}~{\rm yr}$. Notice that for Keplerian rotation, the orbital period at the inner boundary $30r_g$ is $1.03\times 10^3r_g/c=8.03\times 10^{-2}~{\rm yr}$. The magnetic field has the fiducial unit $B_0=2.64\times 10^3~{\rm Gauss}$.

\subsection{Simulation History}
Representative snapshots of the azimuthally-averaged poloidal distribution of density, as
well as the distribution of density and radiation energy density in the equatorial plane, 
are shown in 
Figures~\ref{fig:rhov2Dsnapshots} and \ref{fig:rhomidsnapshots}. It is immediately obvious
that the density distributions undergo significant, non-monotonic variability.
The vertical scale height of the disk goes through cycles of expansion and
contraction, and radially narrow clumps of density
form, dissolve, and reform.  This clump formation process often begins with
the formation of a nonaxisymmetric ($m=2$) density structure which generally transforms into
an axisymmetric ring before eventually diffusing away.  Note that this $m=2$ density pattern is not well-correlated with the radiation energy density in Figure~\ref{fig:rhomidsnapshots}.  At first sight these
variations might appear to be consistent with the predicted behavior of
thermal/viscous instabilities in radiation pressure, electron-scattering
dominated classical alpha disks \citep{SHA76}.  This would predict
runaway heating or cooling and also
an anti-diffusion clumping process \citep{LIG74}.  However, the situation here
is more complex because both the clump formation and the vertical expansion
and contraction are episodes of finite duration and always reverse.

\begin{figure}
    \includegraphics[width=\columnwidth]{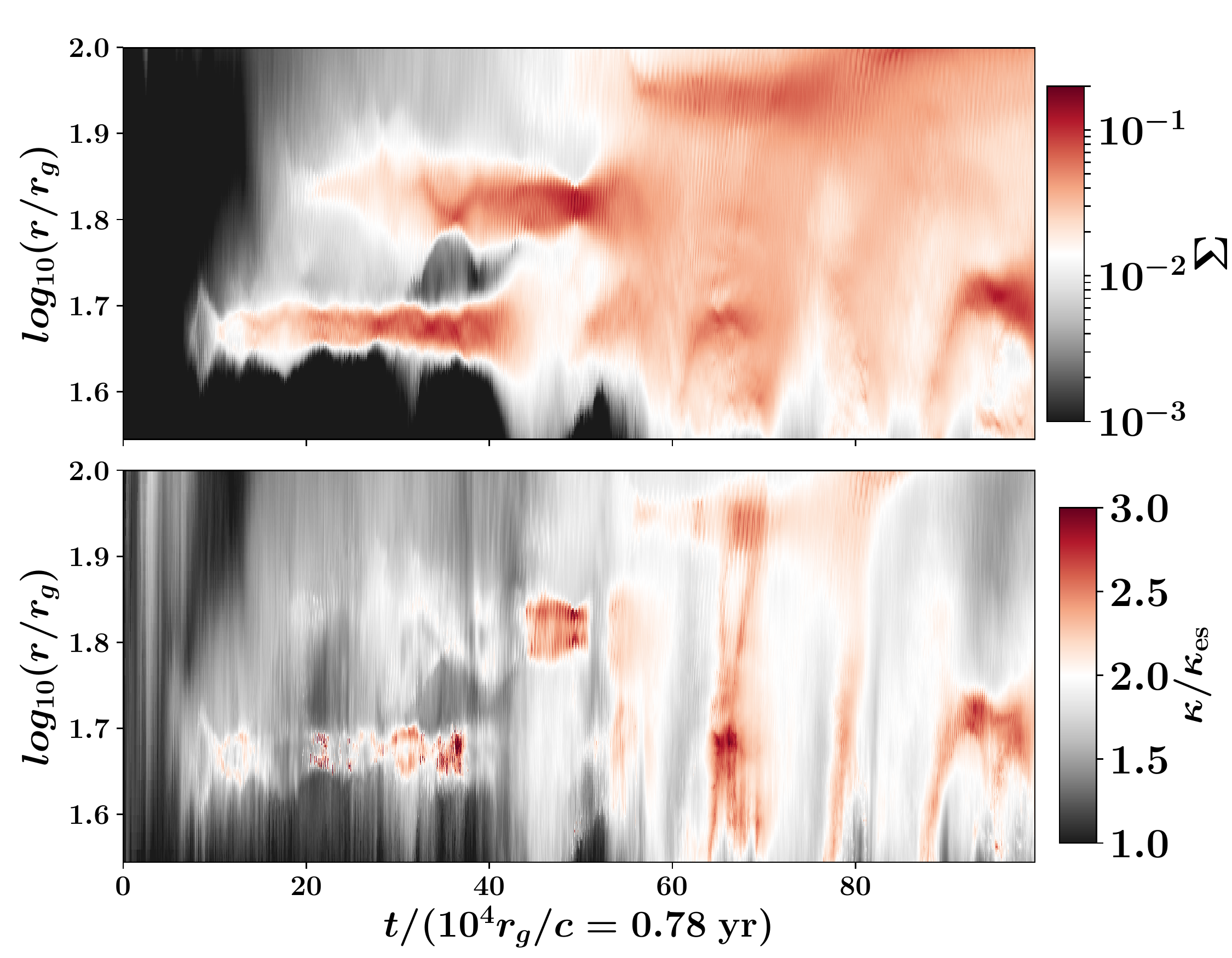}
    \caption{Evolution of shell-averaged surface mass density
(in unit of $\rho_0r_g$)  and ratio of Rosseland opacity to Thomson opacity 
($\kappa_{\rm es}$)
as a function of radius (in units of gravitational radius $r_g$) and time
(in units of $10^4G\mbh/c^3\equiv0.78$~years).
        }
    \label{fig:sigmakappa}
\end{figure}

Figure~\ref{fig:sigmakappa} shows the evolution of shell-integrated surface mass
density and shell-averaged (mass-weighted) opacity as a function of radius and time.
Enhancements in surface density are clearly highly correlated with enhanced
opacity over the electron scattering value.  The iron opacity bump is
playing a critical role in driving the density variability.

\begin{figure}
    \includegraphics[width=\columnwidth]{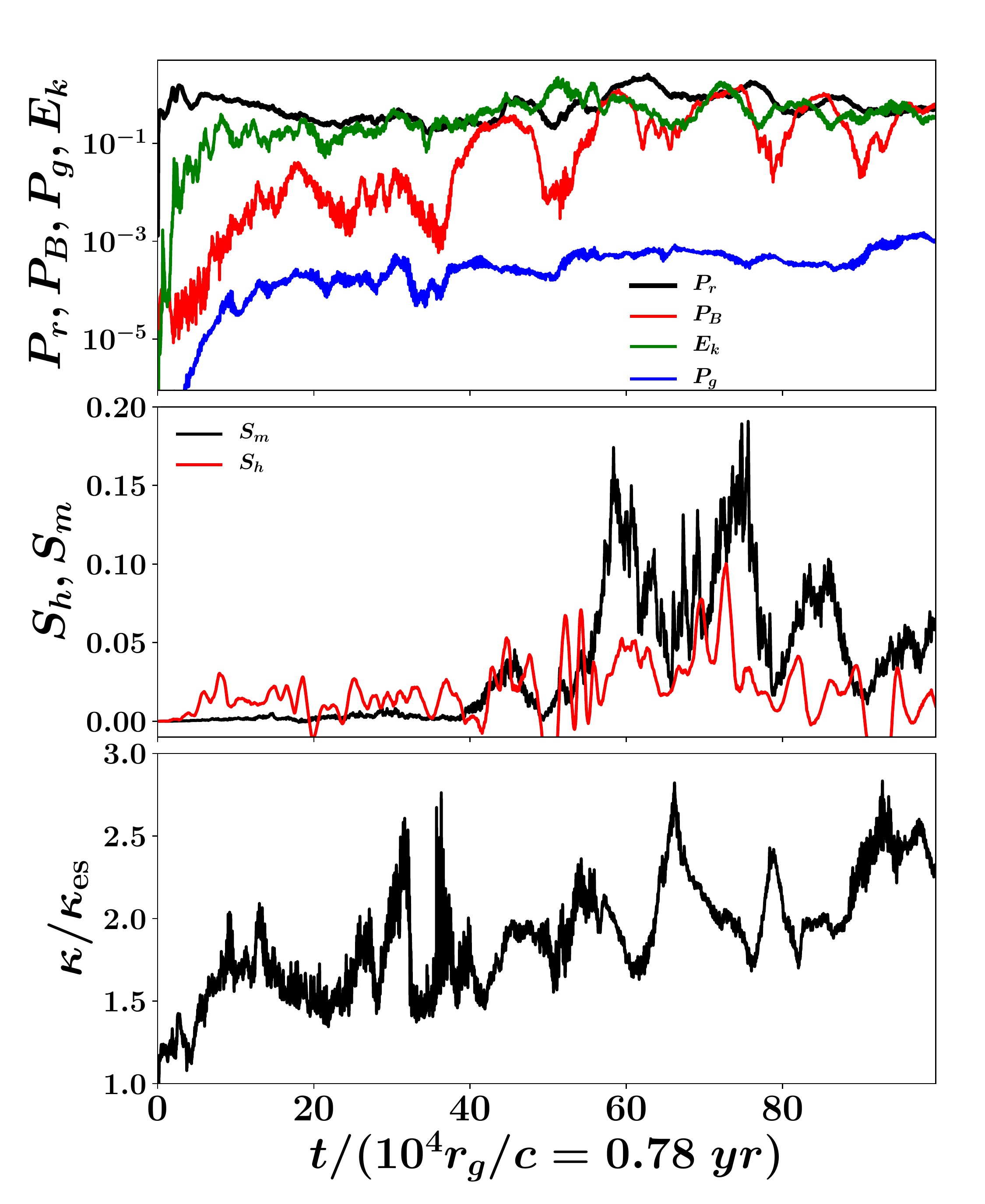}
    \caption{(Top) Evolution of shell-averaged radiation pressure (black), 
    magnetic pressure (red), turbulent kinetic energy density (green) and gas pressure (blue) 
    at radius 50 gravitational radii. These are all scaled with the fiducial pressure 
    unit $P_0=2.77\times 10^5~{\rm dyn/cm}^2$. (Middle) Evolution of shell-averaged 
    Maxwell stress $S_m$ and Reynolds stress  $S_h$ at the same radius. The Reynolds 
    stress is smoothed over the neighboring $100$ data points to reduce noise. Both $S_h$ 
    and $S_m$ are scaled with $P_0$.  (Bottom) Evolution of shell-averaged
    Rosseland mean opacity (scaled with the electron scattering value) at the same radius.
        }
    \label{fig:pressureskappa}
\end{figure}

\begin{figure}[]
    \includegraphics[height=0.9\columnwidth,width=0.9\columnwidth]{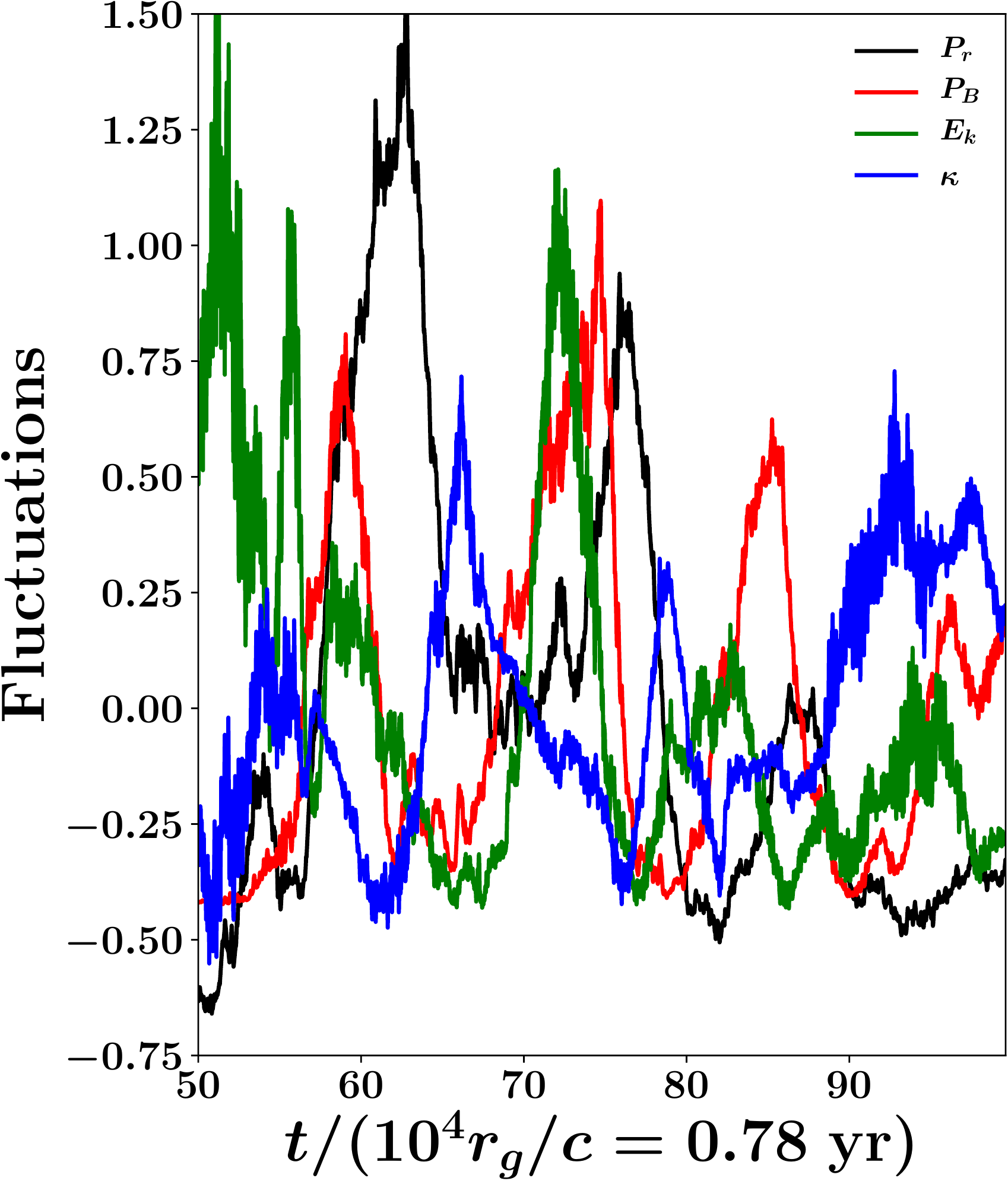}
    \caption{History of the fluctuation components of the shell-averaged radiation 
    pressure (black), magnetic pressure (red), turbulent kinetic energy density (green) 
    and opacity (blue) at $r=50r_g$. The fluctuation components are calculated as the difference 
    between the instantaneous values and the time averaged values between $5\times 10^5r_g/c$ 
    and $10^6 r_g/c$.}
    \label{fig:fluctuation}
\end{figure}

\begin{figure}
    \includegraphics[width=\columnwidth]{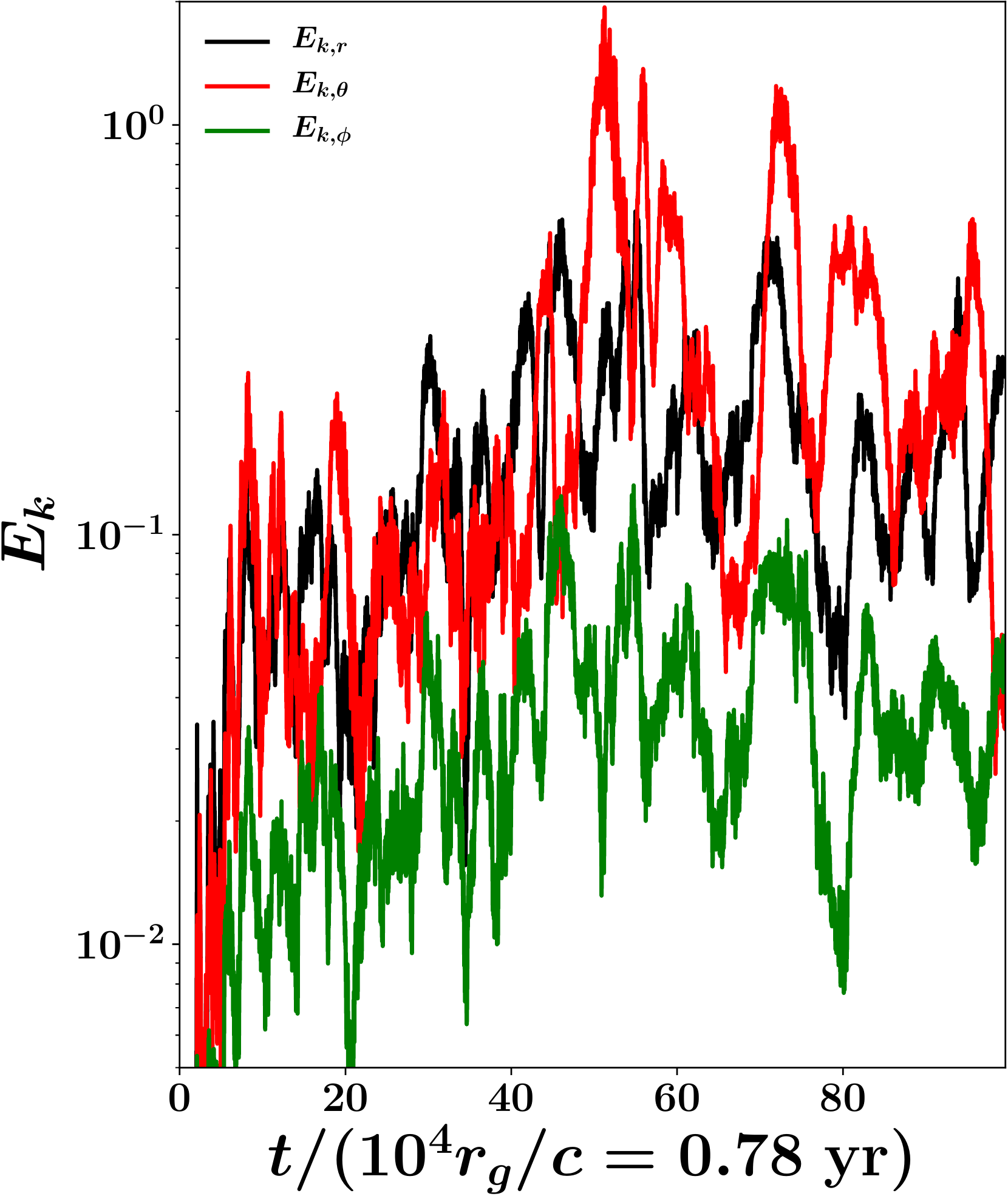}
    \caption{Evolution of shell-averaged turbulent kinetic energy due
to radial, polar, and azimuthal motions at $r=50$~gravitational radii.}
    \label{fig:Ekinhistory}
\end{figure}

\begin{figure}
    \includegraphics[width=\columnwidth]{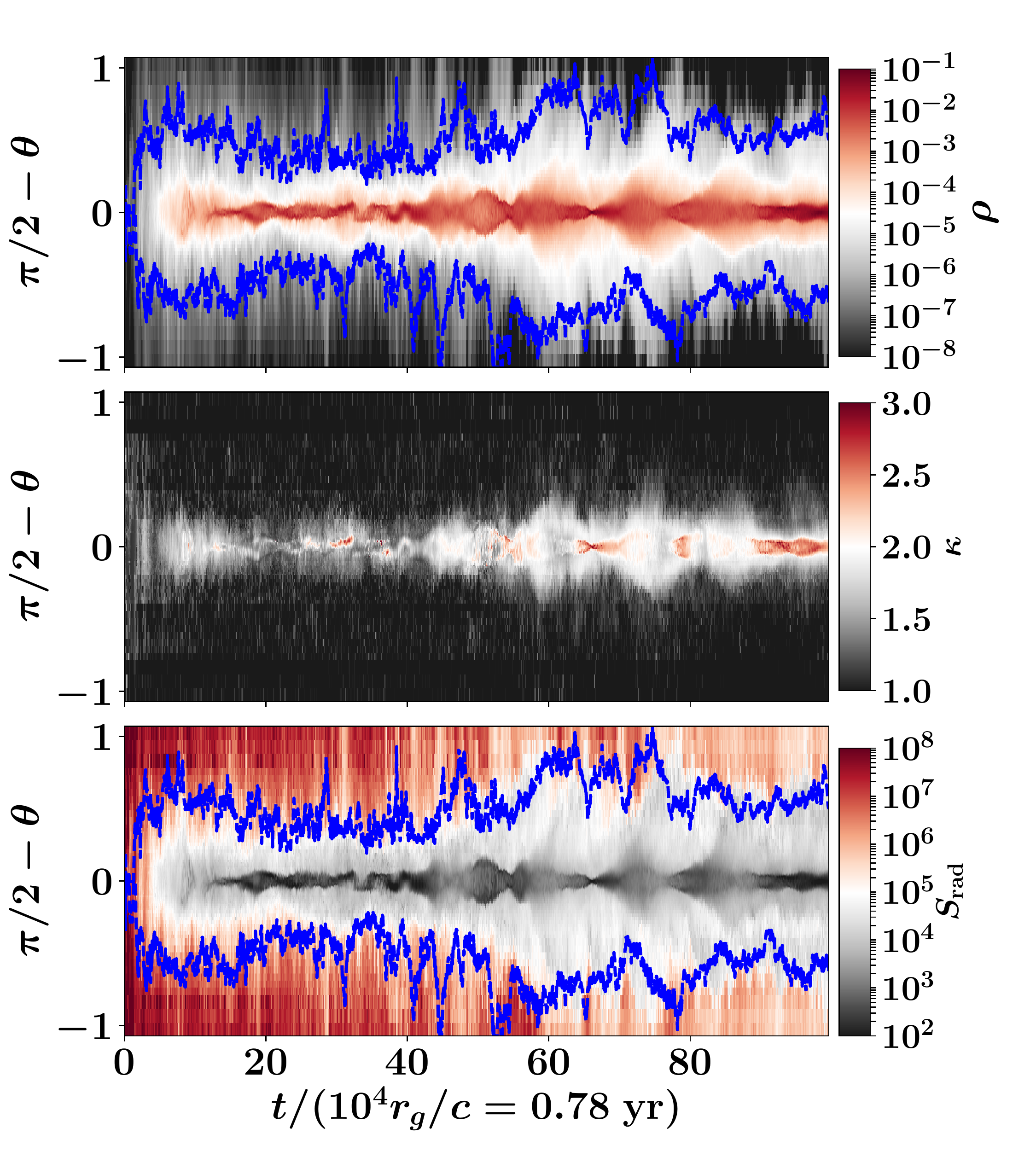}
    \caption{Evolution of azimuthally-averaged density (top), Rosseland
opacity (middle), and radiation entropy per unit mass (bottom) as a function
of height and time at $r=50$~gravitational radii.  The blue dashed lines
show the location of the Rosseland photospheres.
        }
    \label{fig:rhokappaentropyhistory}
\end{figure}

\begin{figure}
    \includegraphics[width=\columnwidth]{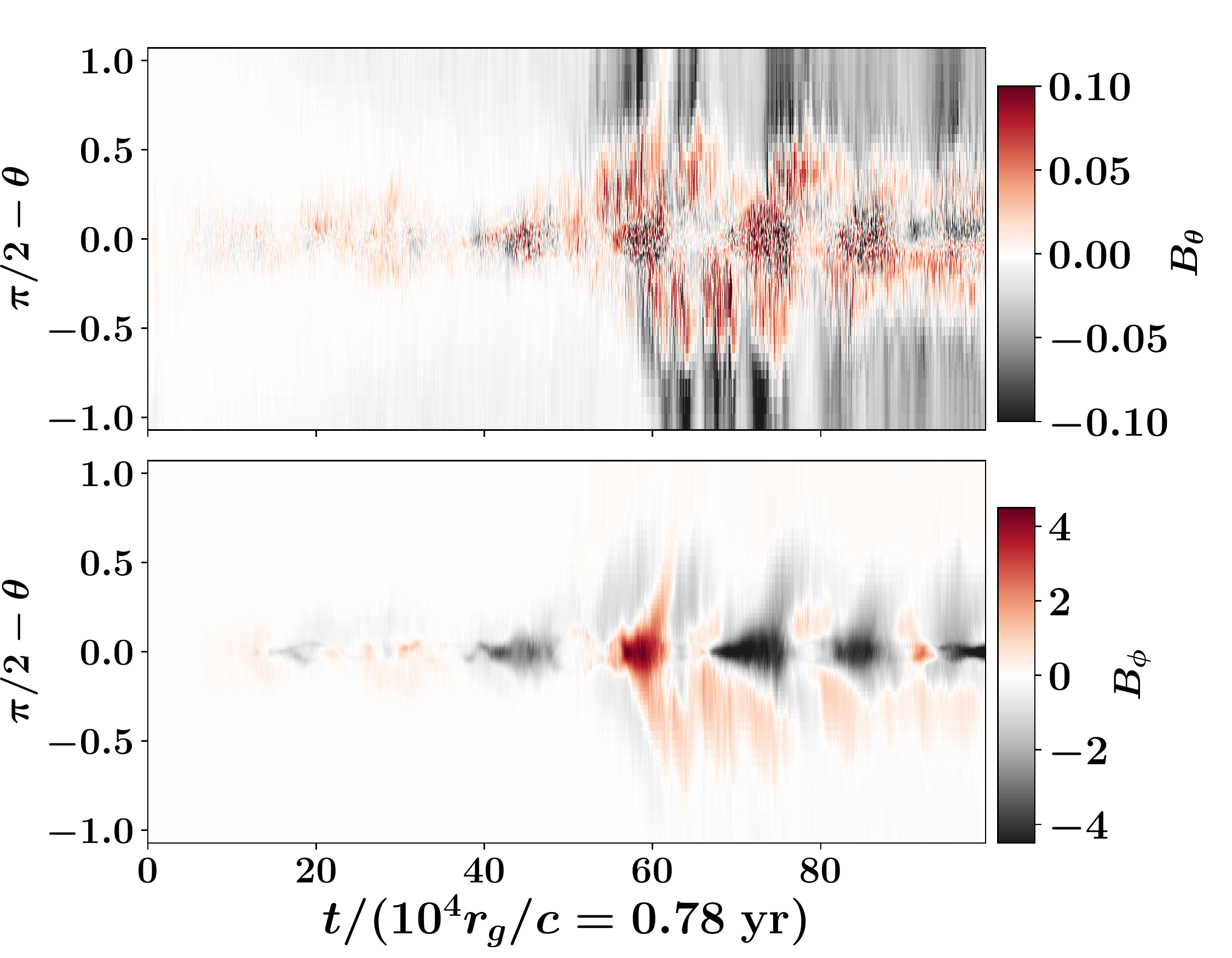}
    \caption{Evolution of azimuthally-averaged polar (top panel) and  azimuthal (bottom panel) 
    magnetic field components as a function of height and time at $r=50$~gravitational radii.
        }
    \label{fig:B2B3history}
\end{figure}

To see why, it is helpful to examine the evolution of various quantities at
a particular radius.  Figure~\ref{fig:pressureskappa} shows the evolution of
various shell-averaged pressures, energy densities and stresses, as well
as the opacity, at $r=50$ gravitational radii.  The turbulent kinetic energy density 
is calculated as $E_k=\rho\left[(v_r-\overline{v_r})^2+(v_{\theta}-\overline{v_{\theta}})^2+(v_{\phi}-\overline{v_{\phi}})^2\right]$/2, 
where $v_r,v_{\theta}$ and $v_{\phi}$ are the three velocity components and 
$\overline{v_r},\overline{v_{\theta}}$ and $\overline{v_{\phi}}$ 
are averaged values (mass weighted) along the azimuthal direction. 
The dominant form of thermal
pressure is radiation pressure, with gas pressure always being completely
negligible.  However, there are also significant, and sometimes dominant,
contributions from turbulent kinetic energy density and magnetic pressure.
The temporal relationship of these quantities is shown more clearly in Figure \ref{fig:fluctuation}.
After $t=50\times10^4r_g/c$, the pressures and energy densities
form a repeating cyclic pattern with large turbulent kinetic energy followed
by magnetic pressure followed by radiation pressure.  These cycles are
clearly correlated with the opacity, with an enhancement
in opacity followed in time by an enhancement in turbulent kinetic energy
density.  Figure~\ref{fig:Ekinhistory} shows how the three components of turbulent
velocity (radial, polar, and azimuthal) contribute to the turbulent kinetic
energy as a function of time.  While MRI turbulence is typically dominated
by radial and azimuthal motions, here the epochs of large turbulent kinetic
energy density are dominated by polar (i.e. vertical in the disk midplane
regions) motions, with radial motions making a secondary contribution.
It is clear that these motions are due to vertical convection driven by
the epochs of enhanced opacity.

\subsection{Opacity Driven Convection}

Figure~\ref{fig:rhokappaentropyhistory} shows the evolution of
azimuthally-averaged density, opacity, and specific entropy at radius
$50 r_g$, but now also as a function of height (represented by
the polar angle $\theta$ near the disk midplane).  This provides more
detail on why the opacity is driving convection:  the creation of unstable
vertical density inversions which are buoyantly unstable (note the drop
in specific entropy as one enters the inversion from below).  The formation
of these inversions is due to the presence of the iron opacity bump.
In an optically thick, radiation pressure supported disk, hydrostatic
equilibrium requires
the temperature to drop vertically away from the midplane.  If the midplane
is on the high temperature side of the iron opacity bump, then opacity
can increase vertically outward, increasing the radiation pressure force
for a given vertical radiation flux.  This can overcome the downward
gravitational force, requiring a large increase in density in order to
have a compensating gas pressure gradient force to maintain hydrostatic
equilibrium.  This is exactly the same thing that happens in one-dimensional
hydrostatic models of massive
star envelopes, and these density inversions trigger convection \citep{JIA15}.
If conditions are optically thick enough that the photon diffusion speed
is much less than the sound speed in the gas alone, then convection is
efficient at transporting heat and generally wipes out the density
inversion.  However, when photon diffusion is fast, convection
is inefficient and the density inversion can survive in a time-averaged
sense \citep{JIA15}.  This is the regime in which low density AGN
accretion disks exist, which is why we can still see the density inversions
in Figure~\ref{fig:rhokappaentropyhistory} in spite of the convective
turbulence.

Very similar behavior to that present in massive star envelopes is therefore
happening here in AGN accretion disks.  However, the situation is even
more interesting here, because the convection is also altering the
MRI dynamo and MRI stresses.  Enhanced convective turbulence can act to
increase the magnetic energy density and to enhance MRI stresses
\citep{HIR14,SCE18}, and this
is evident in Figure~\ref{fig:pressureskappa}:  peaks in turbulent kinetic energy
are always followed by enhanced magnetic energy and enhanced Maxwell stress.
However, this increases turbulent dissipation of accretion power, which
therefore increases the temperature and radiation pressure.  Again, we
are on the high temperature side of the iron opacity bump here, so as the
shell-averaged radiation pressure increases, the shell-averaged opacity
goes down.  (Figure~\ref{fig:rhokappaentropyhistory} shows that these changes
in shell-averaged opacity are reflected in the actual opacities near the disk
midplane during these epochs.)
This then shuts off convection, which reduces MRI turbulent
stress and dissipation, which in turn causes temperature and radiation
pressure to decrease, resulting in opacity increasing again and launching
another cycle of convection.

Maxwell stresses generally dominate angular momentum transport in simulations
of MRI turbulence, but Figure~\ref{fig:pressureskappa} shows that here
Reynolds stresses can also be large, and even at times negative (i.e. driving
inward angular momentum transport).  The negative Reynolds stresses are only
present during the epochs of enhanced turbulent kinetic energy, i.e. when
convection is peaking.  The nonaxisymmetric structures evident in
Figure~\ref{fig:rhomidsnapshots} probably contribute to the epochs of enhanced
%positive - actually maybe negative too, as some of these structures might be leading rather than trailing
Reynolds stress.

Figure~\ref{fig:B2B3history} provides more detail on how convection is affecting
the magnetic field in the midplane regions of the disk.  Azimuthal field
reversals are commonly observed in vertically stratified simulations of
MRI turbulence (the so-called ``butterfly diagram"; \citealt{BRA95,DAV10,HOG16})
and such field reversals are occuring here too.  However,
they are happening on much longer time scales than the usual
$\sim10$ orbital period time scale ($\sim2\times10^4~r_g/c$ at $r=50r_g$).
In fact, the polarity of the field
maintains a consistent sign during epochs of strong convection, with $\sim10$
orbital period field reversals happening only between the convective
epochs, e.g. at $\simeq62\times10^4r_g/c$ in Figure~\ref{fig:B2B3history}.
This is exactly the behavior that is observed in vertically stratified
shearing box simulations of MRI turbulence with convection \citep{COL17}. 
The poloidal component of magnetic field $B_{\theta}$ is also enhanced due to 
convection but with a random sign near the disk midplane.  It was this enhancement of vertical field which was suggested to be the reason behind the enhanced MRI turbulent stresses in strong convection by \citet{HIR14}.  

\begin{figure}
    \includegraphics[width=\columnwidth]{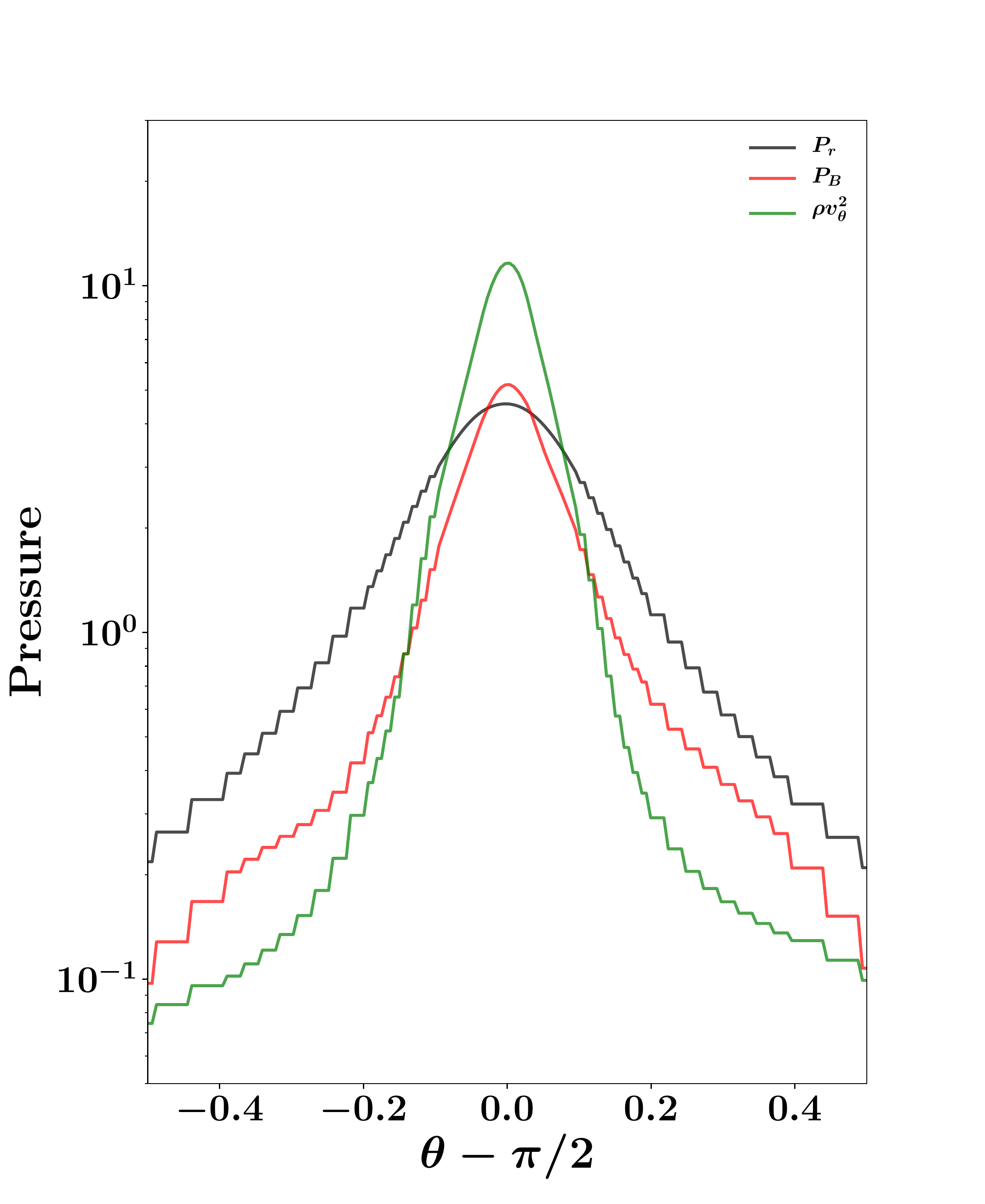}
    \caption{Time and azimuthally averaged vertical profiles of radiation pressure ($P_r$), magnetic pressure ($P_B$) and kinetic term ($\rho v_{\theta}^2$) at radius $50r_g$. The time average is done between $4\times 10^5r_g/c$ and the end of the simulation. All the pressure terms are in unit of $P_0$. The stair step pattern in the profiles is due to prolongation of the data in the region with lower resolutions.  }
    \label{fig:verticalsupport}
\end{figure}

\begin{figure}
    \includegraphics[width=\columnwidth]{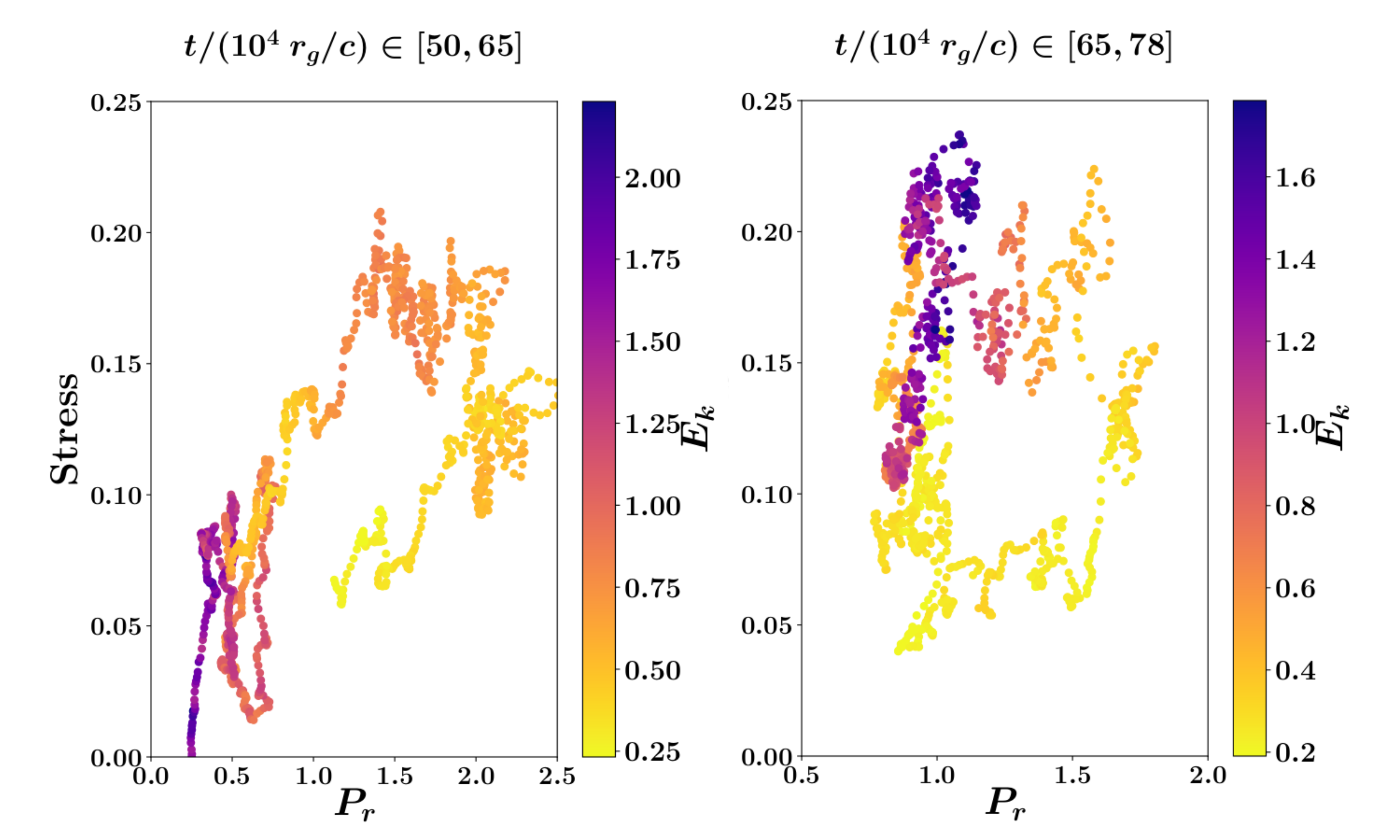}
    \caption{Correlations between the shell averaged total stress (Maxwell plus Reynolds) and radiation pressure  
     at $r=50$~gravitational radii for two oscillation cycles as indicated at the top of each panel. Each data point is color-coded according to the turbulent kinetic energy density. All the variables are scaled with the fiducial pressure unit $P_0$.
        }
    \label{fig:StressPr}
\end{figure}

\subsection{Turbulent Pressure Support in the Disk}
In accretion disks without strong convection, turbulent pressure caused by the MRI turbulence is typically much smaller than the thermal pressure. The disk is usually supported against vertical gravity by gas pressure, radiation pressure or even magnetic pressure \citep{HIR06,BegelmanPringle2007,JIA13,JIA19a}. However, as shown in Figure \ref{fig:pressureskappa}, the turbulent kinetic energy density can be comparable to the radiation pressure in this simulation, which is another characteristic property of radiation pressure dominated convection in the rapid diffusion regime \citep{JIA15,JIA18}. Therefore, the kinetic term $\rho \bv\bv$ in the momentum equation can in principle provide additional support against gravity in the vertical direction. To check this, we plot the time and azimuthally averaged profiles of $P_r$, 
$P_B$ and $\rho v_{\theta}^2$ along the $\theta$ direction at radius $50r_g$ in Figure~\ref{fig:verticalsupport}. The gas pressure is completely negligible here and we neglect it. The gradient of $\rho v_{\theta}^2$ is clearly much larger than the radiation pressure gradient and it balances more than $75\%$ of the gravitational force near the disk midplane in this time-average.
%When convection is off and turbulent pressure is small, the disk midplane region is supported by radiation pressure and the disk scale height is also reduced. {\color{red} I am not convinced that this is happening during non-convective epochs.  E.g. t=62e4 is a non-convective epoch, and yet the photosphere is still very high.  It would help to maybe plot some measure of the disk scale height against turbulent kinetic energy to see if this correlation really holds (which would be interesting and nice!)} 

This provides an alternative or additional explanation as to why the stress is increased when convection is on compared with the suggested mechanism proposed by \cite{HIR14}. Since the typical size of MRI turbulent eddies in the disk is ultimately limited by the disk scale height, the larger the disk scale height, the larger the stress can be. This is also the original argument of the $\alpha$ disk model \citep{SS73}, where the scale height is determined by the thermal pressure and thus the stress is assumed to be proportional to the thermal pressure. Here strong convection-driven turbulent pressure can itself support the disk, allowing a higher stress than we would expect from radiation pressure alone.
%it significantly increases the stress without much increase of the radiation pressure, which explains what we see in Figure \ref{fig:StressPr}.
If we still calculate an effective $\alpha$ as the ratio of stress and radiation pressure, it will be significantly larger when convection is on.

\subsection{Correlations between Stress and Pressure}
As mentioned in the Introduction, a radiation pressure supported accretion disk in the classical $\alpha$ 
disk model is thermally unstable \citep{SHA76}, 
because the total heating rate changes more rapidly with radiation
pressure ($P_r^2$) than the change of the total cooling rate ($P_r$). Although the accretion disk 
structures we find here, as well as the physics we are simulating, are much more complicated than those in the $\alpha$ disk model, it is still interesting 
to check how the stress varies with the radiation pressure while intermittent convection is operating 
in the disk.

The shell averaged total stress as a function of the shell averaged radiation pressure at 
$50r_g$ for two oscillation cycles (within the time intervals $[5,6.5]\times 10^5~r_g/c$ and $[6.5,7.8]\times 10^5~r_g/c$) is shown in Figure \ref{fig:StressPr}. Each data point is color coded with the corresponding turbulent kinetic energy density. When the disk oscillates, the stress and pressure form closed loops in this plot. When convection is active, as indicated by the large turbulent kinetic energy density,
stress increases rapidly while $P_r$ increases more slowly.
%from $\approx 0.05$ to 
%$\approx 0.23$ while $P_r$ increases by roughly a factor of two.
This heating then reduces the opacity, turning off convection and decreasing $E_k$.  The stress then decreases while $P_r$ continues to increase further, presumably because of the dissipation of the convective turbulent kinetic energy and magnetic energy.
%$E_k$ decreases, stress decreases while $P_r$ increases by another factor of two.
Finally, both stress and $P_r$ decrease at roughly the same rate. This confirms that when convection is on, stress follows turbulent kinetic energy density closely. The heating rate increases more rapidly than the change of the cooling rate and that is why the disk heats up. When convection is off, the heating and cooling rate have roughly the same dependence on the radiation pressure.  This is perhaps why the disk does not undergo a runaway collapse during the phase when it cools down, which is similar to what \cite{JIA16} found.

\begin{figure}
    \includegraphics[width=\columnwidth]{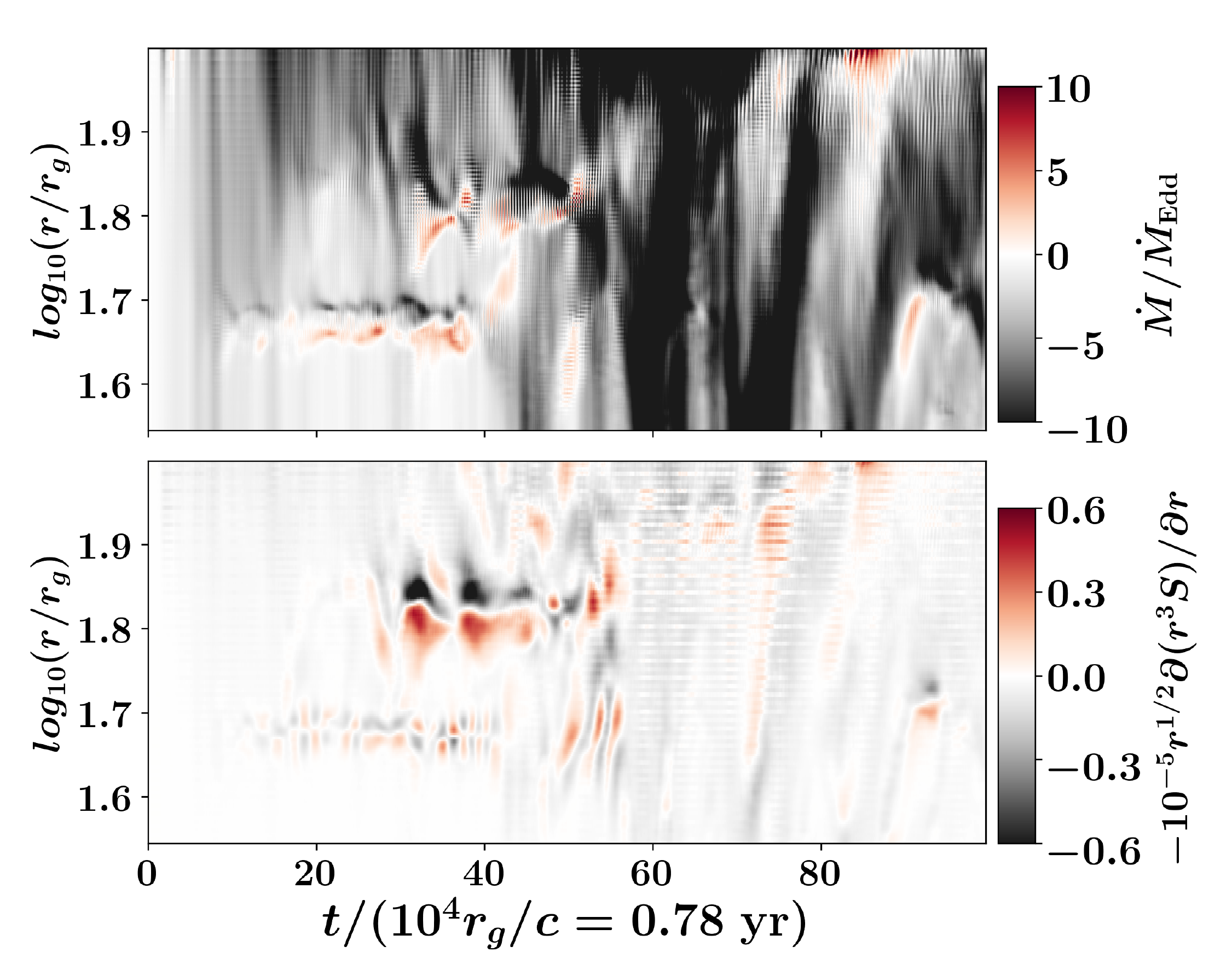}
    \caption{Space-time diagrams of the local mass accretion rate $\dot{M}$ (top panel, in unit of 
    the Eddington mass accretion rate $\Medd$), and derivatives of the total stress 
    $-r^{1/2}\partial \left(r^3 S\right)/\partial r$. Negative and positive values of $\dot{M}$ mean inward 
    and outward accretion correspondingly. 
    Both $S$ and $\dot{M}$ are smoothed over 
    the neighboring 100 data points in time to reduce the noise. The Eddington accretion rate is defined 
    as $10\Ledd/c^2$ with $\Ledd$ to be the Eddington luminosity.} 
    \label{fig:mdotstress}
\end{figure}

\begin{figure}
    \includegraphics[width=\columnwidth]{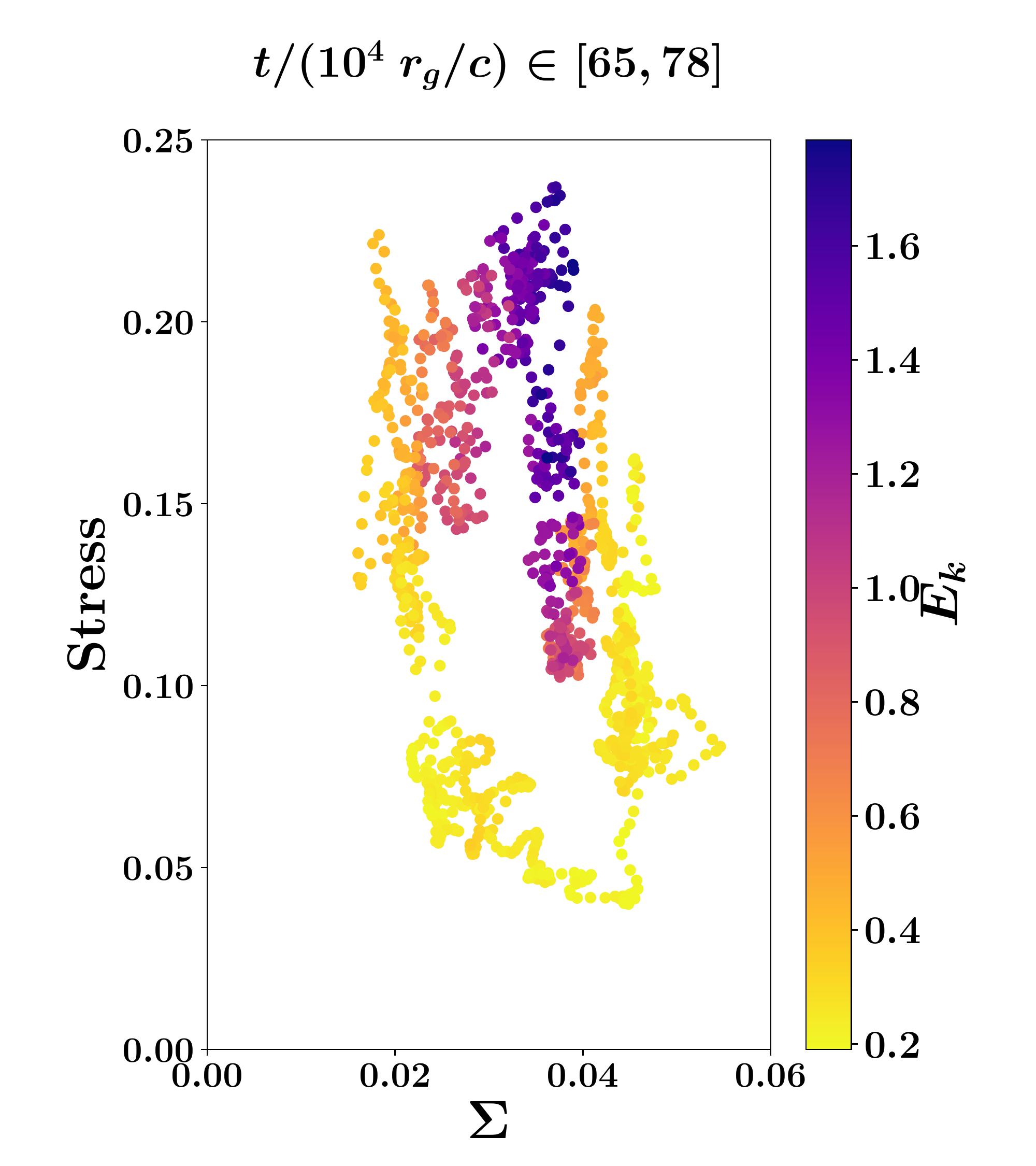}
    \caption{Correlations between surface density $\Sigma$ and shell averaged total 
    stress at $r=50$~gravitational radii for the convective cycle between $[65,78]\times 10^4~r_g/c$. 
    Each data point is color coded with the corresponding turbulent kinetic energy density.
        }
    \label{fig:stresssigma}
\end{figure}

\subsection{Radial Mass Diffusion and Clumping}

With the assumption that angular momentum transport is dominated by local turbulent stresses, the vertically averaged equations of mass and angular momentum conservation can be used to write an equation for surface density evolution \citep{BAL99}:
\begin{equation}
    \frac{\partial\Sigma}{\partial t}=-\frac{1}{2\pi R}\frac{\partial\dot{M}}{\partial R},
    \label{eq:dsigmadt}
\end{equation}
where $R$ is the cylindrical polar coordinate radius and the accretion rate (assumed negative for inflow) is given by
\begin{equation}
    \dot{M}=-\frac{2\pi}{\ell^\prime}\frac{\partial}{\partial R}(R^2W_{R\phi}).
    \label{eq:mdot}
\end{equation}
Here $\ell^\prime$ is the radial specific angular momentum gradient and $W_{R\phi}$ is the vertically integrated turbulent stress.  In viscous or alpha-disk theory, these two equations can be combined to give a radial mass diffusion equation \citep{LYN74,LIG74}, but we choose not to do that as it is the radial gradients in stress that most clearly drive the clumping of surface mass density observed in our simulation.

That this is so may be seen in Figure \ref{fig:mdotstress}.  The upper panel shows a space-time plot of the shell-averaged mass accretion rate $\dot{M}$, and it is clear that radial gradients of this quantity match very well the clumping pattern observed in the surface density evolution in Figure~\ref{fig:sigmakappa}, in accordance with equation (\ref{eq:dsigmadt}).  Of course, this
had to be true as it merely tests mass conservation in {\sf Athena++}.
%\footnote{A perfect match would require us to use the mass fluxes at cell faces computed by the the Riemann solver in {\sf Athena++}, %which we have not bothered to do here.  {\color{red} Is this correct, Yan-Fei?}}

Less trivial is equation~(\ref{eq:mdot}), which relies on the assumption that all angular momentum transport is done through local turbulent stresses rather than non-local processes
(e.g. the spiral waves that are evident in Figure~\ref{fig:rhomidsnapshots}).  If this is true, then a plot of $-r^{1/2}(\partial/\partial r)(r^3S)$, where $S$ is the shell-averaged Maxwell plus turbulent Reynolds stress, should resemble the pattern in accretion rate.  This is plotted in the lower panel of Figure \ref{fig:mdotstress}, and does indeed approximately match the accretion rate behavior shown in the upper panel.

It is therefore radial gradients in the turbulent stresses that are largely responsible for the clumping.
%and these radial gradients are present because the stresses are enhanced by opacity-driven convection (bottom panel of Figure~\ref{fig:sigmakappa}).
These radial gradients can be
strong enough that mass can actually sometimes diffuse outward, as is evident in the upper panel of Figure \ref{fig:mdotstress}.  Note from
the bottom panel of this Figure that clumping
is occurring because there is a radially local
deficit of stress.  Even though convection enhances the stress overall, high opacity is
actually anticorrelated in time with stress in the convective cycles shown in Figure~\ref{fig:pressureskappa}, and this produces
the local deficit.  As convection kicks in and the stress is enhanced, and the opacity drops, the clump diffuses away.

Note that the derivation of the pure ``viscous" instability associated with electron-scattering and radiation pressure dominated classical alpha-disk accretion models relies on an assumption of local thermal equilibrium in order to derive an inverse relationship between stress and surface mass density \citep{LIG74}.  This results in an effective negative diffusion coefficient in the radial mass diffusion equation that results from combining equations (\ref{eq:dsigmadt}) and (\ref{eq:mdot})
\citep{PRI81}.  This analysis can be generalized to include departures from thermal equilibrium \citep{SHA76}, but the coupling with varying opacity in the convective cycles clearly makes things far more complicated here.  We have attempted to analyze the local behavior of stress as a function of surface density, just as we did with stress as a
function of radiation pressure in Figure~\ref{fig:StressPr}.  There are epochs where loops in such a diagram form and there is some evidence of stress being inversely proportional to surface density when there is no convection present, e.g. the bottom of the loop in Figure \ref{fig:stresssigma} which shows the cycle between $[65,78]\times 10^4r_g/c$ at $r=50r_g$.  However, this inverse trend is broken by the onset of iron opacity-driven convection, and this behavior is not always generic.  It is therefore unclear that such
a classical analysis is appropriate in the
presence of these complex convective cycles.

\subsection{Resolution}

To check how well the MRI turbulence is resolved during different phases of the oscillation 
cycles in the simulation, we calculate the ratios between the wavelength of the fastest growing 
MRI mode and the cell sizes along the polar and azimuthal directions, i.e. the
quality factors $Q_{\theta}$ and $Q_{\phi}$ \citep{HAW11,SOR12}. These are widely used in non-radiative ideal MHD simulations and indicate that MRI turbulence is fully resolved when 
$Q_{\phi}\gtrsim 25, Q_{\theta}\gtrsim 6$ or both of them are larger than 10. 
Following \cite{JIA19a} (section 3.1), we also use them as a check for our radiation MHD simulations. 
For the first representative snapshot shown in Figure \ref{fig:rhov2Dsnapshots}, $Q_{\phi}$ stays around $40$ near the disk midplane for radii smaller than $\approx 55r_g$ and then drops to 
$11$ inside the high density clump. Similarly, $Q_{\theta}$ stays around $7$ until reaching the high 
density clump, where it drops to $2$. At time $t=8\times 10^5r_g/c$ when the disk expands, $Q_{\phi}$ varies
from 30 to 100 over the whole radial range from $30r_g$ to $100r_g$, while $Q_{\theta}$ varies from $20$ 
to $\approx 3$. When the disk collapses at $t=6.5\times 10^5r_g/c$, $Q_{\phi}$ varies from 30 inside $45r_g$ to 
$10$ from $45r_g$ to $\approx 100r_g$. The averaged $Q_{\theta}$ varies from 6 to 3 over the same radial range. This suggests that MRI turbulence is reasonably well-resolved in this simulation and 
we have the worst resolution when the disk collapses, which is not surprising. Fortunately, the accretion does not stop during the collapsing phase as the opacity-driven oscillation cycle continues. 

\begin{figure}
    \includegraphics[width=\columnwidth]{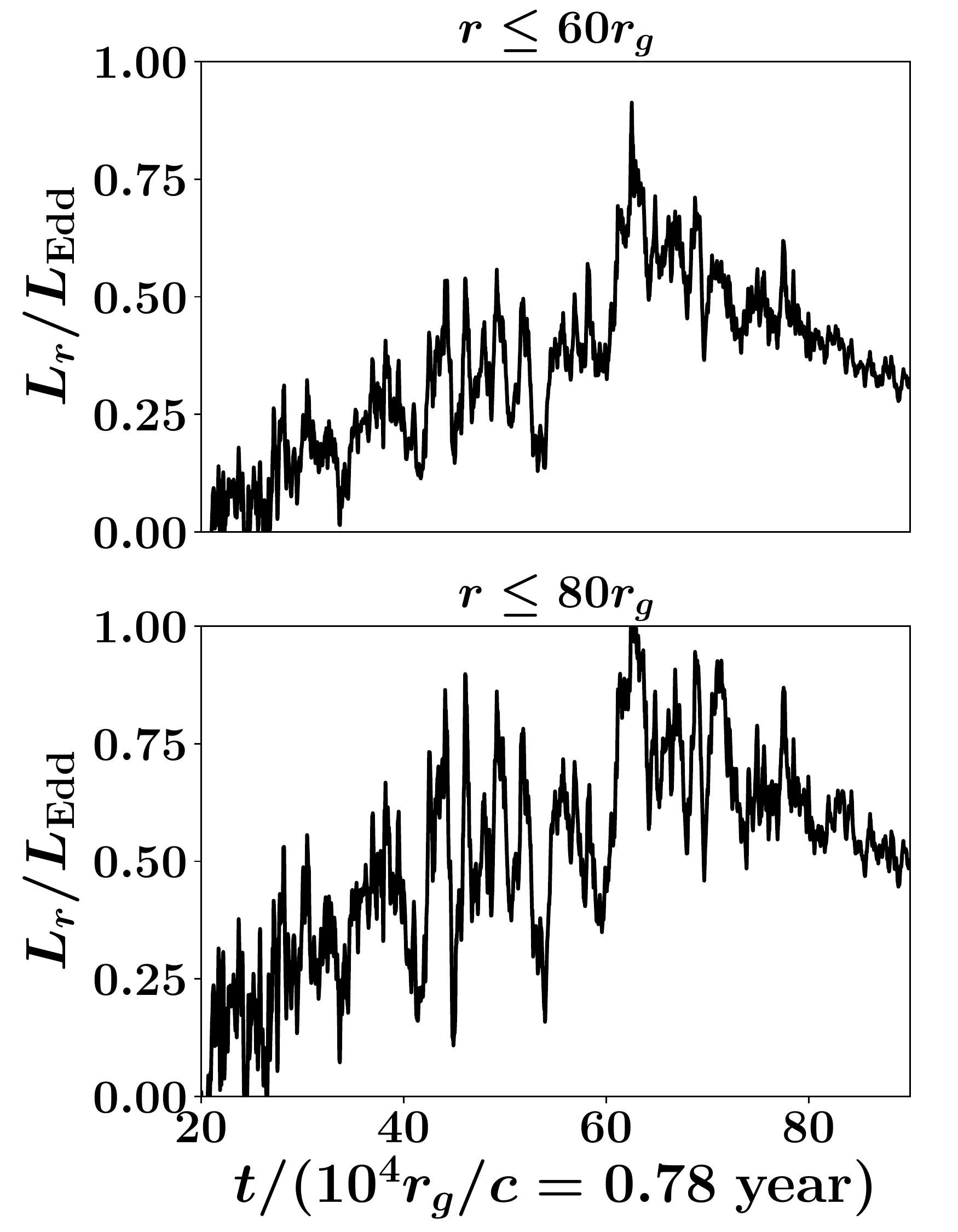}
    \caption{History of the total luminosity $L_r$ (scaled with the Eddington luminosity $L_{\rm Edd}$) coming from the disk. The top panel shows the luminosity if we only include the disk inside  $60r_g$ while the bottom panel shows the luminosity inside 
    $80r_g$.
        }
    \label{fig:Lumhistory}
\end{figure}

\subsection{Lightcurve Variability}

The disk oscillation cycles driven by the opacity also cause the total luminosity coming from
the photosphere to vary significantly with time. The total luminosities in the simulation emerging from radii
inside $60r_g$ and $80r_g$, respectively, are shown in Figure \ref{fig:Lumhistory}. 
Normal MRI turbulence without convection can cause the luminosity to vary by a factor of $\sim 2$ over the local thermal time scale. Smaller amplitude variability over the local dynamical time scale can also show up in the luminosity when the optical 
depth across the disk is low enough (see Figure 1 of \citealt{JIA19a}). However, with convection driven oscillations in the disk, the luminosity can vary by a factor of $\approx 3- 6$ over the local thermal time scale, which is roughly a few years in this radial range.

\section{Discussion}

The time scale of luminosity variations depicted in Figure~\ref{fig:Lumhistory} are remarkably consonant with those observed in changing look quasars, and we also see amplitudes of variation by as much as a factor of four.  In addition, the variations in scale height of the photosphere depicted in Figure~\ref{fig:rhokappaentropyhistory} occur on comparable time scales and will effect the ability of this region of the disk to intercept and reprocess radiation from the very inner disk, as well as shadow larger radii of the disk.

The typical variability time scale driven by convection is determined by the local thermal time scale 
at the radial range where the iron opacity bump is located. Since the thermal time scale is roughly related to the local dynamical 
time scale by $1/\alpha$, the variability time scale will change when the iron opacity bump moves to different radii for different black hole masses and mass accretion rates. For a fixed mass accretion rate in Eddington units, the disk temperature will typically decrease with increasing black hole mass at fixed $r/r_g$. This means the iron opacity bump, which is roughly at a fixed temperature around $1.8\times 10^5$ K, will move closer to the black hole. At the same time, the local dynamical time scale will also increase linearly with black hole mass for a fixed $r/r_g$. The combination of the two effects makes the thermal time scale at the location where the iron opacity is located very insensitive to the black hole mass. 
In fact, for the classical inner accretion disk in the \citet{SS73} model, the disk midplane temperature scales as $T\propto(r/r_g)^{-3/8}M^{-1/4}$, and is even independent of the accretion rate.  This then gives a thermal time scale for fixed midplane temperature independent of both black hole mass and accretion rate, though it is very sensitive to the temperature of the iron opacity bump as well as the alpha parameter: $t_{\rm thermal}\propto \alpha^{-2}T^{-4}$.  Real accretion disks will be more complicated than this classic model, as we have tried to demonstrate here.  Even so, this suggests that the rapid luminosity variation time scales that we have found in this one simulation may be quite common across many AGNs. However, the amplitude of variability driven by this mechanism may depend on the mass accretion rate and black hole mass when the iron opacity bump moves to different radii from the central black hole. When it is further away from the black hole and when the surface density is smaller due to lower accretion rate, we expect the variability amplitude will get smaller. This is perhaps one reason why changing look AGNs only make up 10 percent of the quasar population.

AGN variability has been widely parameterized with stochastic models such as the Auto-Regressive Moving Average (ARMA) approach \citep{Kellyetal2009,Kellyetal2014,Morenoetal2019}. Such modeling provides valuable information on the physical properties of the disk such as the typical timescales associated with the variability. Although the lightcurve from our simulation is still very preliminary and only extends over a short period of time, we have tried ARMA modeling of the luminosity as well as the history of magnetic energy density from the simulation using the {\emph{statsmodels}} package \citep{seabold2010statsmodels}. A lower order ARMA(2,1) model can fit the simulation data very well. Both luminosity and magnetic energy density have the rise timescale of variability around one year. But the variability amplitude for magnetic energy density is larger. This preliminary fitting demonstrates that the simulation data shares some similar stochastic properties to observed AGN lightcurves and it might therefore be possible to constrain the physical properties of the observed system. However, more detailed comparison is beyond the scope of this paper and  will be studied in the future. 

Note from Figure~{\ref{opacity}} that all the effects we have been discussing in this paper are driven by small (factors of $3-5$) enhancements in the Rosseland opacity over the electron scattering value in the iron opacity bump.  This is of course due to the huge dominance of radiation pressure over gas pressure in quasar accretion disks, so that even small variations in opacity can have enormous consequences.  We assumed solar abundances for our simulation, whereas AGN accretion disks are likely to have significantly supersolar metallicities (\citealt{Fieldsetal2007,Aravetal2007}). The variability amplitude driven by convection will likely increase with larger metallicity for a given black hole mass and accretion rate, which can be compared with observed properties of AGNs \citep{JIA16}. %It would be interesting to try and correlate observed variability behavior with measured iron abundances. {\color{red} This last is actually a point that was made by Shane Davis at the Edinburgh conference, so I am nervous about making it without having him as a co-author.}
%{\color{red} We have already made a similar point in the discussion of the 2016 paper. I am not sure this is a significant new idea. But I have slightly modified the words.}

One dimensional models of cooler annuli further out in the disk can themselves exhibit density inversions \citep{HUB00}.  Those inversions are due to ionization transitions of hydrogen and helium, as the models of \citet{HUB00} did not include any metals.  It is possible that the convective effects we have explored here also happen in these regions of the disk, though if anything they are likely to be more dramatic, as is the case in massive stars when hydrogen and helium opacity effects come into play \citep{JIA18}.

The convective cycles we have observed here share many similarities to those observed in stratified shearing box simulations of local patches of disks in cataclysmic variables \citep{HIR14,COL17,SCE18,COL18} and protostellar disks \citep{HIR15}.  This includes the intermittency of convection\footnote{Persistent convection has only been observed in local simulations of helium-dominated accretion disks by \citet{COL18}.} and the enhanced stresses and persistent magnetic polarity during epochs of convection.
%All those previous simulations were done with rather narrow radial box domains, which can affect the properties of MRI turbulence \citep{SIM12}.
This is the first time that these effects have been confirmed in a global simulation.  However, this is in a very different regime of radiation pressure dominated flows, and the enhanced vertical support caused by the convective turbulence itself may be a contributing factor to the enhanced stresses.
This possible new mechanism for convection-driven enhancement of turbulent stresses cannot explain the enhanced stresses observed in shearing box simulations of gas-pressure dominated disks \citep{HIR14, HIR15, SCE18, COL18}.  There the turbulent kinetic energies were always much less than the thermal pressure, and did not contribute significantly to vertical support of the disk.  In fact, convective epochs typically had smaller vertical scale heights than radiative epochs.  The large turbulent kinetic energies which are present here are due to radiation pressure dominated convection in a regime where the photon diffusion speed is larger than the gas sound speed.

In fact, we also observe such supersonic convection in this regime in massive star envelopes \citep{JIA18}.  In that case, the energy source for these supersonic motions is the flow of heat from the core of the star.   The turbulent velocity is much smaller than the radiation sound speed deep in the star and can become comparable to the local radiation sound speed near the photosphere. This is possible because the pressure scale height is $\approx 20\%$ of the stellar radius and the size of the turbulent eddies are comparable to the background radiation pressure scale height. A buoyant fluid element accelerated in the deeper regions can reach a velocity comparable to the local radiation sound speed when it moves to a larger distance. For the case of accretion disks, the energy driving convection ultimately arises from the accretion power. As we have shown in Figure \ref{fig:mdotstress}, the effects of convection are also not strictly local in radius, as radial mass motions result from the
change of angular momentum transporting stresses.  The convection is therefore able to
tap into the free energy stored in differential
rotation, and reach supersonic velocities.
%The energy source for these supersonic motions in convection is directly coming from the thermal energy, which is radiation in the radiation pressure dominated flow. Gas pressure is irrelevant and the turbulent kinetic energy can exceed the gas internal energy. In the case of massive star envelops \citep{JIA18}, the radiation energy density is coming from the core of the star. The turbulent velocity is much smaller than the radiation sound speed deep in the star and can become comparable to the local radiation sound speed near the photosphere. This is possible because the pressure scale height is $\approx 20\%$ of the stellar radius and the turbulent eddies can feel the background radiation pressure gradient. The fluid element accelerated in the deeper region can reach a velocity comparable to the local radiation sound speed when they move to a larger distance. For the case of accretion disks, the radiation energy is ultimately coming from the accretion power. As we have shown in Figure \ref{fig:mdotstress}, convection is also not strictly local in radius. The turbulent velocities triggered by the opacity bump reach the peak values at a larger radius. The turbulent kinetic energy can also become comparable or larger than the local radiation energy density. }

We have not explained the formation of the $m=2$ nonaxisymmetric density structures that are evident in Figure~\ref{fig:rhomidsnapshots}.  One possibility is that when a surface density peak is formed in a localized radial range as shown in Figure \ref{fig:sigmakappa}, the disk is potentially subject to the Rossby wave instability, which has been widely studied for protoplanetary disks 
\citep{Lovelaceetal1999,Lietal2001,Lyra2012}. This instability can create vortices and excites high frequency waves and even spiral 
density structures in the disk. \cite{Lovelaceetal1999} show that the sufficient condition for the Rossby wave instability is that the 
the inverse potential vorticity multiplied by the entropy  function $S$, which is $S^{2/\Gamma}\Sigma/\left(\bfnabla\times \bv
\right)_{z}$, has a local maximum as a function of radius in the disk. Here $\Gamma$ is the adiabatic index. It is currently unclear how this instability criterion can be generalized to the radiation pressure dominated regime with realistic 3D structures as in our simulation. Nevertheless, we checked this criterion using $\Gamma=4/3$ and the radiation entropy per unit mass.  Indeed, this function does show a local maximum at the location where the high density clump is located. In fact, this function already shows local extrema before the density clumps and spiral patterns can be clearly seen in the disk as shown in Figure \ref{fig:rhomidsnapshots} and \ref{fig:sigmakappa}. We will leave the detailed study of Rossby wave instability in AGN disks for future studies. But this suggests that it is one possible mechanism to explain the nonaxisymmetric structures we have found in the simulation.

A major caveat of our simulation is that it is so expensive that we can only afford to run the simulation for a few thermal 
timescales for the inner $\sim 60r_g$. The time-averaged mass accretion rate is not a constant as a function of radius, which is necessary if the disk is in steady state. This could either be because the simulation time is not long enough, or a steady state disk is simply not possible when convection driven oscillation are operating. We hope to investigate this further with future calculations of longer duration. 

%\section{Conclusions}
%{\color{red} Do we need a conclusion section?}

\section*{Acknowledgements}

Resources supporting this work were provided by the NASA High-End Computing (HEC) Program through the NASA Advanced Supercomputing (NAS) Division at Ames Research Center.
OB acknowledges the support of a 2019 JILA visting fellowship which enabled very
helpful conversations with Jason Dexter on changing look quasars and with
Ben Brown, Evan Anders, and Imogen Cresswell on the physics of supersonic
convection.  We also thank Matt Coleman, Shigenobu Hirose, Yue Shen and all the members from the Horizon Collaboration for useful discussions.  The Center for Computational Astrophysics at the Flatiron Institute 
is supported by the Simons Foundation. This research was also supported in part by the National Science Foundation under Grant No. NSF PHY-1748958

\bibliographystyle{aasjournal}
\bibliography{citations}

\end{CJK*}

\end{document}